\documentclass{elsart-accepted}

\usepackage{natbib}
\bibpunct{(}{)}{;}{a}{}{,}

\usepackage{graphicx}

\usepackage[latin1]{inputenc}

\usepackage{amssymb}

\usepackage{url}

\journal{Deep-Sea Research II}
\date{2005-12-21}

\begin{document}

\begin{frontmatter}


 \title{
       Reply to comment on `A simple
model for the short-time evolution of near-surface current and temperature
profiles'
       }
\author[labelADJ]{Alastair D.~Jenkins\corauthref{cor1}},
 \ead{alastair.jenkins@bjerknes.uib.no}
 \ead[url]{\url{http://www.gfi.uib.no/~jenkins}}
\corauth[cor1]{Corresponding author. Tel.: +47-55 58 2632; fax: +47-55 58 9883.}
\author[labelBW]{Brian Ward}
\address[labelADJ]{Bjerknes Centre for Climate Research, Geophysical Institute,
Allégaten 70, 5007 Bergen, Norway}
\address[labelBW]{Woods Hole Oceanographic Institution,
       Woods Hole, Massachusetts, U.S.A.
}




\begin{abstract}
This is our response to a comment by Walter Eifler on our paper `A simple
model\break for the short-time evolution of near-surface current and
temperature profiles'\break (arXiv:physics/0503186, accepted for publication in
{\em Deep-Sea Research II\/}).  Although Eifler raises genuine issues regarding
our model's validity and applicability, we are nevertheless of the opinion that
it is of value for the short-term evolution of the upper-ocean profiles of
current and temperature.  The fact that the effective eddy viscosity tends to
infinity for infinite time under a steady wind stress may not be surprising.
It can be interpreted as a vertical shift of the eddy viscosity profile and an
increase in the size of the dominant turbulent eddies under the assumed
conditions of small stratification and infinite water depth.
\end{abstract}


\end{frontmatter}


Eifler~(\citeyear{EiflerW:dsr-2005-xxx}) criticises our
model~\citep[][referred to as {\em JW\/}]{JenkinsAD-WardB:blm-2005-xxx} 
for a number of reasons:
\begin{enumerate}
  \item\label{reasonA} If our model is run with a
step-function wind and wind stress applied at time $t=0$, the velocity field
$u(z,t)$ becomes
\begin{equation}
u(z,t) = \lambda U \exp\left(-{\lambda U\over{u^*}^2}{z\over t}\right),
\label{eq-one}
\end{equation}
and satisfies
\begin{equation}
{\partial u \over \partial t} = {{u^*}^2z\over\lambda U}{\partial^2
u\over\partial z^2}, \label{eqA}
\end{equation}
where $z$ is the downward-pointing vertical coordinate, $\lambda U$ is the
interface velocity, $U$ is the wind speed, and $u^*$ is the waterside friction
velocity. In {\em JW\/} we state that the quantity
\begin{equation}
\varepsilon_{\it app} = {{u^*}^2z\over\lambda U}
\end{equation}
plays a role similar to that of an eddy viscosity.  {\em Eifler\/} disagrees
with our interpretation, using a stricter definition of eddy viscosity
$\varepsilon$, in our one-dimensional case it is defined using the equation
\begin{equation}
{\partial u \over \partial t} = {\partial\over\partial z}\left(\varepsilon{\partial
u\over\partial z}\right),
\end{equation}
from which he deduces from~(\ref{eqA}) that
\begin{equation}
\varepsilon = {{u^*}^2\over\lambda U}z + \left({{u^*}^2\over\lambda U}\right)^2 t,
\label{eq-varepsilon}
\end{equation}
so that the vertical momentum flux `has to be sustained by an eddy viscosity
tending towards an infinite value'.
  \item\label{reasonB} {\em Eifler\/} computes the time dependence of the
mixed-layer depth (defined for the {\em JW\/} model as the depth where the
local vertical momentum flux is a specified fraction of the surface momentum
flux), and compares it with an `empirically-tested standard formulation'.  The
mixed-layer depth $z_{\it ml}$ of the {\em JW\/} model obeys
\begin{equation}
z_{\it ml} = {A^{T=x\%}\over 16}u^* t, \label{eq-zml}
\end{equation}
where $A^{T=x\%}$ is defined by {\em Eifler\/}'s Eq.~10.  The
empirically-tested standard formulation gives a mixed-layer depth $D_{\it ml}$
which obeys
\begin{equation}
D_{\it ml} = 10.5 u^* t^{1/2}, \label{eq-Dml}
\end{equation}
the units of $D_{\it ml}$, $u^*$, and $t$ being metres, m\,s$^{-1}$, and
seconds, respectively.
The different time dependence of the mixed layer depths in Eqs.~\ref{eq-zml}
and~\ref{eq-Dml} is stated to be a reason against the validity of the {\em JW\/}
model.
  \item\label{reasonC} The derived numerical thermal model of {\em JW\/} shows
in Fig.~3 (of {\em JW\/}) 
a mixed layer depth which develops linearly in time, and
the agreement with observations is not particularly good in this case, giving
further evidence against the usefulness of the {\em JW\/} model.
\end{enumerate}

Although {\em Eifler\/} raises genuine issues which question the validity and
applicability of the {\em JW\/} model, we are nevertheless of the opinion that
it has a degree of usefulness: indeed, {\em Eifler\/} does state in his
conclusion that {\em JW\/} `have provided a promising analytical solution of
the  
vertical transport equation $\ldots$ with very interesting 
mathematical properties'.  

Regarding {\em Eifler\/}'s Reason~\ref{reasonC} above, we concede that the
model (not surprisingly, given its simplicity) does not always give good
agreement with the measured data, although the particular reasons for this may
also include horizontal advection effects.  Of course, perfect dynamical
agreement cannot in any case be obtained, since the {\em JW\/} model assumes
zero Coriolis force, neutral stratification, and infinite water depth.  

Reason~\ref{reasonA}, that the eddy viscosity $\varepsilon$ tends to infinity for
infinite time under a steady applied wind stress, is not such a serious
objection as it first appears.  Equation~\ref{eq-varepsilon} may be
re-written as\begin{equation}
\varepsilon = {{u^*}^2\over\lambda U}\left[z + \left({{u^*}^2\over\lambda
U}\right) t\right],
\end{equation}
from which we see that the graph of $\varepsilon$ versus $z$ shifts
upwards at a velocity $\left[{{u^*}^2/(\lambda
U})\right]$.  This result may also be interpreted as an increase in the
length scale of the dominant turbulent eddies which impact the surface
(and other depths).  The increase of $\varepsilon$ to infinity at
infinite time is a consequence of the depth being infinite.  The fact
that the near-surface values of the eddy viscosity are no longer close
to linearly proportional to depth is in fact reminiscent
of the breaking-wave enhanced values of near-surface eddy viscosity
shown in Figs.~3.18, 3.19, and 7.16 of \cite{BurchardH:atmmw-2002}.

Reason~\ref{reasonB}, that the {\em JW\/} linear $t$ dependence of 
mixed layer depth $z_{\it ml}$ is in disagreement with the
empirically-tested $t^{1/2}$ dependence of mixed layer depth $D_{\it
ml}$, is rather interesting.  Equation~\ref{eq-Dml} corresponds to
Eq.~6.1 in \cite{BurchardH:atmmw-2002}, which was suggested by
\cite{PriceJF:jfm-1979-509}.  The constant 10.5 in~(\ref{eq-Dml}) is in
fact $1.05N_0^{-1/2}$, with $N_0$, the initial Brunt-Väisälä
frequency, being set to~10$^{-2}$\,s$^{-1}$.  For an initially
linearly-stratified water column, or, in fact, for a water column with 
any stably-stratified density profile, subjected to a 
constant input of turbulent
energy at the surface for mixing, the result $D_{\it ml} \propto
t^{1/2}$ is a consequence of the requirement that the potential energy
of the water column increases at a constant rate due to mixing.
Similarly, the result $z_{\it ml}\propto t$ is a consequence of a
constant rate of momentum input to the water column.  
If the Coriolis force due to the Earth's rotation is taken into
account, $z_{\it ml}$ can be expected not to increase indefinitely,
as the wind-induced current would be restricted to a boundary layer
of Ekman depth typically of the order of magnitude of $u^*/f$
\citep{MadsenOS:jpo-1977-248}.  Further study is required to resolve these
different boundary layer time dependences, and to evaluate the
joint effects of stratification and rotation within the simple {\em JW\/}
model framework.

\subsection*{Acknowledgements}

The work was funded by Research Council of
Norway projects 127872/720 and 155923/700.  
This is Publication No.~000 from the Bjerknes Centre
for Climate Research.

\end{document}